\documentclass{article}
\usepackage[utf8]{inputenc}
\usepackage{authblk}
\usepackage{float}
\usepackage{tikz}
\usepackage{soul}
\usepackage{multicol}
\usepackage{smartdiagram}
\usepackage{caption}
\usepackage{arev}
\usetikzlibrary{mindmap}
\usepackage{pdflscape}  
\usepackage{csvsimple}
\usepackage{pdflscape}
\usepackage{hyperref}
\usepackage{subcaption}
\usepackage{graphicx}
\usepackage{rotating}
\usepackage{comment}
\usepackage{multirow}
\usepackage{lineno}
\usepackage{array}
\usepackage{lscape}
\usepackage[backend=biber,style=vancouver,sorting=none]{biblatex}
\usepackage{setspace}
\DeclareUnicodeCharacter{0301}{\'{o}}
\DeclareUnicodeCharacter{202F}{ }
\usepackage{amssymb}
\usepackage[a4paper, margin=1in]{geometry}
\usepackage{times}

\makeatletter
\renewcommand{\maketitle}{\bgroup\setlength{\parindent}{0pt}
\begin{flushleft}
  \textbf{\@title}

    \@author
    \end{flushleft}\egroup
    }
    \makeatother

    \providecommand{\keywords}[1]
    {
      \textbf{\textit{Keywords---}} #1
      }

      \font\myfont=cmr12 at 17pt

      \title{\myfont{Explainable Artificial Intelligence For The Detection and Characterisation of Stage B Heart Failure}}

      \date{}
      \author[1,2]{Ahmed M Salih}
      \author[1,2]{Emer Brady}
      \author[1,2]{Ranjit Arnold}
      \author[1,2]{Gaurav Gulsin}
      \author[2,3]{Huiyu Zhou}
      \author[1,2]{Anvesha Singh}
      \author[1,2]{Gerry McCann}

      \affil[1]{Division of Cardiovascular Sciences, University of Leicester, Leicester LE1 7RH, UK}
      \affil[2]{British Heart Foundation Leicester Centre of Research Excellence, University of Leicester and the National Institute for Health and Social Care(NIHR) Biomedical Research Centre Leicester, Glenfield Hospital, Leicester, UK}
      \affil[3]{School of Computing and Mathematical Sciences, University of Leicester, Leicester LE1 7RH, UK}

\begin{document}
      \maketitle
      \thispagestyle{empty} 

      \noindent
      \begin{abstract}
      \noindent
       \textbf{Background:} Stage B heart failure is characterised by asymptomatic structural or functional cardiac abnormalities. Identifying individuals at this stage of diseases is clinically important, as early detection may enable the use of targeted interventions to prevent progression to symptomatic heart failure. Explainable artificial intelligence (XAI) may facilitate early detection, transparent risk stratification, and appropriate selection of clinically actionable intervention to prevent disease progression.\\
      \textbf{Objective:} To review the use of XAI methods in detecting and characterising stage B heart failure.\\
      \textbf{Methods:} A literature search of Web of Science, Scopus, and PubMed was conducted on 27 March 2026. Studies were included if they applied AI with XAI techniques to stage B heart failure. After screening, 20 studies were included. Data on modalities, outcomes, demographic reporting, and XAI methods were extracted and synthesised.\\
      \textbf{Results:} SHAP was the most commonly used explainable AI method, followed by LIME, saliency maps, and Grad-CAM; however, XAI adoption remained inconsistent, with several studies relying on limited or ad hoc interpretability approaches. Importantly, none of the studies generated or compared explanations across sex or ethnic subgroups, despite evidence of subgroup-specific differences in disease burden. Furthermore, evaluation of XAI outputs was often lacking: some studies did not assess explanations at all, while others relied solely on literature-based comparisons, which may introduce bias and subjectivity. Together, these limitations indicate that explainability was not systematically validated or leveraged to support fair and generalisable clinical inference..\\ 
      \textbf{Conclusion:} XAI shows promise for improving transparency in Stage B heart failure identification, but current implementations remain limited. Major gaps include insufficient consideration of sex and ethnicity, lack of subgroup-specific explainability analyses, inconsistent XAI evaluation, and limited external validation, all of which constrain generalisability and clinical adoption.

      \end{abstract}
      \newpage
\section{Introduction}
Heart failure (HF) remains a major global health challenge, affecting more than 60 million people worldwide and contributing substantially to morbidity, mortality, and healthcare expenditure~\cite{chopra2025icardio}. Although HF is typically diagnosed at symptomatic stages, mounting evidence demonstrates that the disease process begins much earlier, with subtle structural or functional cardiac abnormalities that precede overt clinical manifestations. This early stage is referred to as stage B heart failure (SBHF), which is asymptomatic and represents a critical window during which targeted interventions may prevent or delay progression to symptomatic HF~\cite{cowley2026challenges}. Detecting individuals at this stage therefore is clinically relevant, yet conventional diagnostic approaches, including electrocardiography, echocardiography, and biomarker assessment, may lack sufficient sensitivity or interpretability to reliably identify these early abnormalities.\\
Artificial intelligence (AI) has emerged as a powerful tool in cardiovascular medicine, offering enhanced pattern recognition and predictive capabilities beyond traditional statistical methods~\cite{romiti2020artificial}. AI models have demonstrated promise in identifying SBHF from diverse data sources, including imaging, physiological signals, and laboratory biomarkers. However, the clinical translation of AI remains limited by the “black-box” nature of many algorithms, which obscures the reasoning behind model predictions and restricts clinician trust, regulatory acceptance, and integration into clinical decision-making.\\
Explainable AI (XAI) seeks to address this challenge by providing transparent, interpretable insights into model behaviour. Techniques such as SHapley Additive exPlanations (SHAP), Local Interpretable Model-Agnostic Explanations (LIME), saliency maps, integrated gradients, and Gradient-weighted Class Activation Mapping (Grad-CAM) offer mechanisms to visualize feature importance, highlight influential imaging regions, and reveal underlying decision pathways~\cite{salih2024review}. In the context of SBHF, XAI has the potential not only to improve clinicians' confidence in AI-derived predictions but also to accelerate discovery of novel disease mechanisms, risk markers, and prognostic indicators.\\
This review therefore aims to synthesise current evidence on the use of XAI techniques in identifying SBHF. We summarize the AI and XAI methods employed across studies, highlight trends in data sources and clinical outcomes, and discuss key challenges and opportunities for advancing interpretable AI in the SBHF.
\section{Stage B Heart Failure}
\begin{figure}[H]
    \centering
    \includegraphics[width=\linewidth]{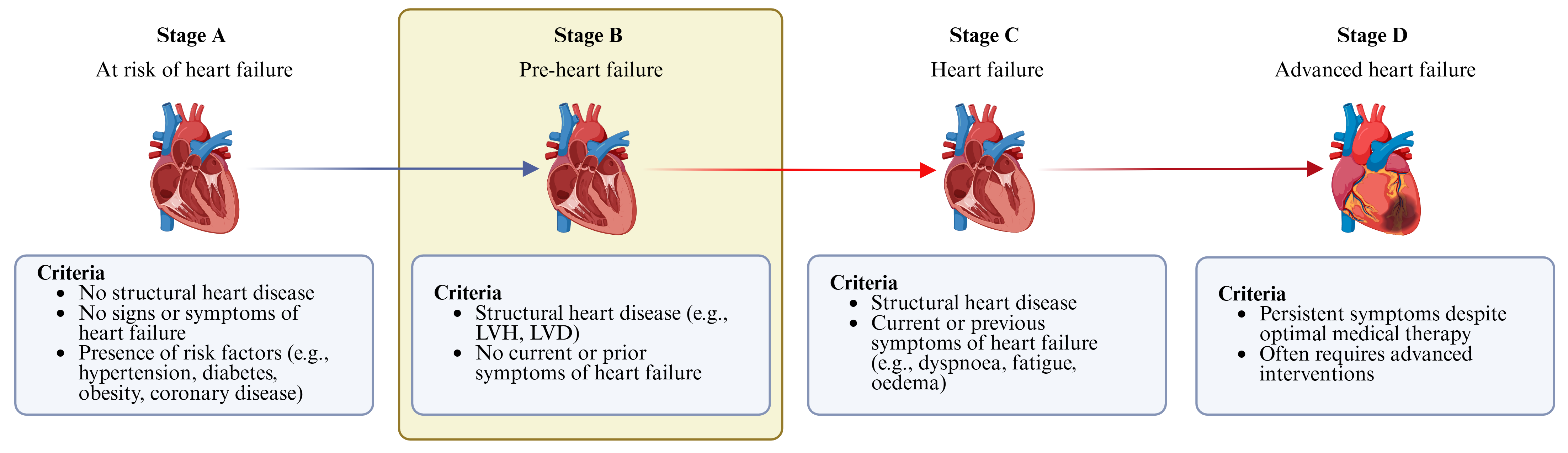}
    \caption{Heart failure stages. LVH: left ventricular hypertrophy; LVD: left ventricular dysfunction.}
    \label{XAI_stages}
\end{figure}

HF is defined by the 2022 American Heart Association (AHA)/American College of Cardiology (ACC) / Heart Failure Society of America (HFSA) guideline as a clinical syndrome resulting from structural and/or functional cardiac abnormalities that impair the heart's ability to meet the metabolic demands of the body. Typical symptoms are breathlessness, fatigue with or without signs of fluid retention. The guideline emphasizes HF as a progressive condition, occurring across a continuum from risk factors to advanced symptomatic disease, where early recognition and prevention are crucial~\cite{heidenreich20222022}.\\
Figure~\ref{XAI_stages} illustrates the stages of HF based on the 2022 AHA/ACC/HFSA guideline~\cite{heidenreich20222022} defining four distinct phases in HF development. Stage A, termed “At risk for HF,” includes individuals without structural heart disease or symptoms but who possess risk factors known to predispose them to HF, such as hypertension, diabetes, obesity, or a family history of cardiomyopathy. Importantly, these individuals have no imaging or biomarker evidence of cardiac injury, and interventions focus on aggressive risk factor modification.
\noindent
SBHF, now defined as “Pre-heart failure,” includes people who remain asymptomatic but demonstrate structural heart disease or elevated filling pressures, or have persistent elevations in biomarkers indicative of myocardial stress or injury~\cite{bayes2024pre}. SBHF is primarily identified by cardiac imaging revealing structural or functional abnormalities such as left ventricular hypertrophy or reduced ejection fraction, and may be supported by elevated circulating biomarkers, particularly brain natriuretic peptide. These structural abnormalities reflect early cardiac remodelling and place individuals at increased risk for progression to symptomatic HF. Prevention strategies in SBHF emphasize guideline directed medical therapies to halt or slow disease progression.\\
Across the literature, SBHF is described using heterogeneous terminology, including asymptomatic structural heart disease, left ventricular dysfunction~\cite{zeng2022multimodal}, subclinical left ventricular dysfunction, left ventricular hypertrophy~\cite{soto2022multimodal}, left ventricular remodelling, and preclinical heart failure. These terms generally refer to the same clinical construct of structural or functional cardiac abnormalities in the absence of overt heart failure symptoms. For clarity and consistency, this manuscript adopts the AHA/ACC/HFSA classification and refers to this stage uniformly as SBHF.\\
Previous studies showed that SBHF might progress over time to symptomatic heart failure with potential to regress to stage A. A subset of individuals with SBHF were observed to regress to stage A in a study conducted as part of the Olmsted County Heart Function Study~\cite{young2021progression}. This reversibility highlights SBHF as a critical window for the application of AI methods to identify high-risk individuals early and enable timely, targeted interventions before irreversible disease progression occurs.
\section{Explainable AI}
\begin{figure}[H]
    \centering
    \includegraphics[width=\linewidth]{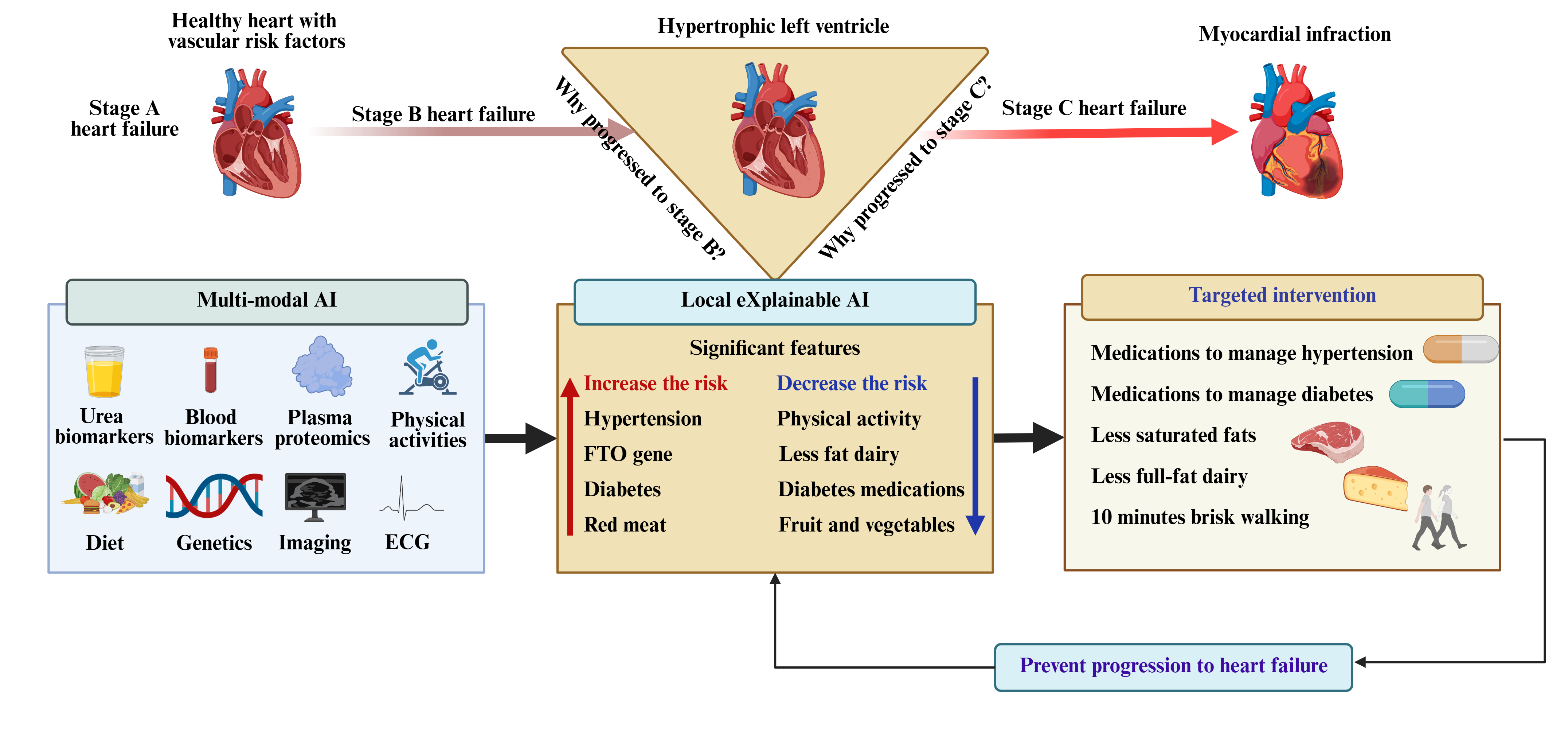}
    \caption{XAI integration with heart failure prohression.}
    \label{XAI_HF}
\end{figure}

XAI refers to a set of methods that help end-users (e.g., clinicians) understand why an AI model has made a particular prediction. While AI can analyze large volumes of clinical, imaging, and biomarker data far beyond what the human eye can detect, clinicians often hesitate to rely on these systems because many operate as "black boxes" producing outputs without showing their reasoning. XAI aims to bridge this gap by providing clear, interpretable information that supports clinical decision-making rather than replacing it~\cite{salih2024characterizing}.\\
XAI methods are commonly categorised into post hoc and ante hoc (intrinsic) approaches. Post hoc methods are applied after the development and training of an AI model and require additional techniques to interpret how the model arrives at its predictions. These approaches aim to provide insights into otherwise opaque models without modifying their internal structure. In contrast, ante hoc methods are inherently interpretable, meaning that the model itself is designed to be transparent, and its decision-making process can be directly understood without the need for supplementary explanation techniques. Another important taxonomy of XAI distinguishes between local and global explanations. Local explanations focus on interpreting the prediction for an individual instance, providing insight into why a specific outcome was generated for a particular patient or data point. This is particularly useful in clinical settings where patient-specific decisions are required. In contrast, global explanations aim to describe the overall behaviour of the model across the entire dataset, identifying general patterns, feature importance, and decision logic that apply to the population as a whole.\\ 
Similar to AI models, XAI methods also require systematic evaluation to ensure that the generated explanations are reliable, meaningful, and clinically useful. Broadly, three main evaluation approaches have been proposed.
First, application-grounded evaluation involves assessing XAI outputs within a real-world context by domain experts, such as cardiologists in cardiovascular applications. In this approach, explanations are qualitatively evaluated based on criteria such as completeness (whether the explanation captures all relevant factors), plausibility (whether it aligns with clinical knowledge), and complexity (whether it is understandable and actionable for clinicians). Second, proxy or functionality-grounded evaluation focuses on quantitative assessment without direct involvement of domain experts. This approach relies on predefined metrics, axioms, or statistical measures to evaluate explanation quality. Examples include methods such as RemOve And Retrain (ROAR), selectivity, normalized movement rate, and sensitivity analysis, which aim to quantify how well the explanation reflects the model's true decision-making process. Finally, literature-grounded evaluation compares the explanations generated by XAI methods with established findings in the scientific literature. In this case, the validity of the explanation is assessed based on its consistency with previously reported clinical or physiological knowledge.
Each of these evaluation approaches has distinct strengths and limitations, and their appropriate use depends on the intended application and clinical context. A detailed discussion of these methods, including their advantages and drawbacks, can be found in~\cite{salih2024review}.\\
In cardiology, the need for XAI is particularly important. For example, as shown in Figure~\ref{XAI_HF}, when an AI model predicts that a patient may have SBHF, clinicians naturally want to know which features the model considered and whether those features align with known pathophysiology. XAI methods may highlight the ECG leads or waveforms that contributed most strongly to a prediction, allowing the clinician to verify that the model focused on clinically meaningful regions, such as voltage criteria or repolarisation patterns. Similarly, in echocardiography, XAI tools may indicate whether the model paid attention to wall thickness, chamber size, or subtle changes in myocardial deformation when suggesting early structural abnormalities. Moreover, multi-modal XAI has the ability to identify the diet type and medications that are associated with outcome either positively or negatively. The indicated increases/decreases in features and the suggested interventions in Figure~\ref{XAI_HF} are hypothetical and intended solely for illustrative purposes.
\noindent
Another practical benefit of XAI is that it may support risk communication with patients. For instance, an AI model estimating a patient's likelihood of progressing from SBHF to Stage C heart failure may be accompanied by an explanation showing that elevated blood pressure, increased left ventricle mass, and persistent strain abnormalities were the main contributors. This type of transparency can help clinicians guide discussions around lifestyle modification, medication intensification, or closer follow-up.\\
As healthcare increasingly incorporates AI-based decision tools, XAI will be essential to ensure trust, safety, and clinical usability. By revealing how predictions are generated, XAI allows cardiologists to integrate AI insights with their own clinical expertise, ultimately improving early detection and personalized prevention of heart failure.
\section{Methods}
The review included original research studies that employed AI methods and incorporated XAI or interpretability techniques to investigate SBHF. A literature search was performed in Web of Science, Scopus, and PubMed on 27 March 2026. The search strategy combined terms related to early or preclinical cardiac dysfunction (e.g., Stage B heart failure, asymptomatic left ventricular dysfunction, subclinical LV dysfunction, left ventricular remodelling) with terms representing explainable AI methods (e.g., XAI, interpretability, SHAP, LIME, Grad-CAM, integrated gradients).
These example keywords reflect the two core conceptual domains of interest: early cardiac structural/functional changes and explainable AI techniques. The full set of search terms used for database queries is provided in Supplementary Table~\ref{keywords_}. Boolean operators (AND/OR) were applied to combine terms from both domains to identify studies implementing XAI methods within AI models for detecting or characterizing SBHF. Studies were excluded if they did not apply AI, applied AI without the use of XAI or interpretability approaches, lacked accessible full text, or were non-original publications such as review articles, commentaries, editorials, or conference abstracts without complete papers.\\
For each included study, we extracted key information including the data modality, the use of external validation, incorporation of demographic variables (e.g., sex and ethnicity) for model adjustment, predicted outcomes, implemented XAI methods, and the approaches used to evaluate these explanations. In addition, for studies reporting the use of demographic data or external validation, we examined whether the explanations generated by the XAI methods differed when applied to external datasets or across demographic subgroups relative to the original training data.
\section{Results}
\begin{figure}[H]
    \centering
    \includegraphics[width=0.5\linewidth]{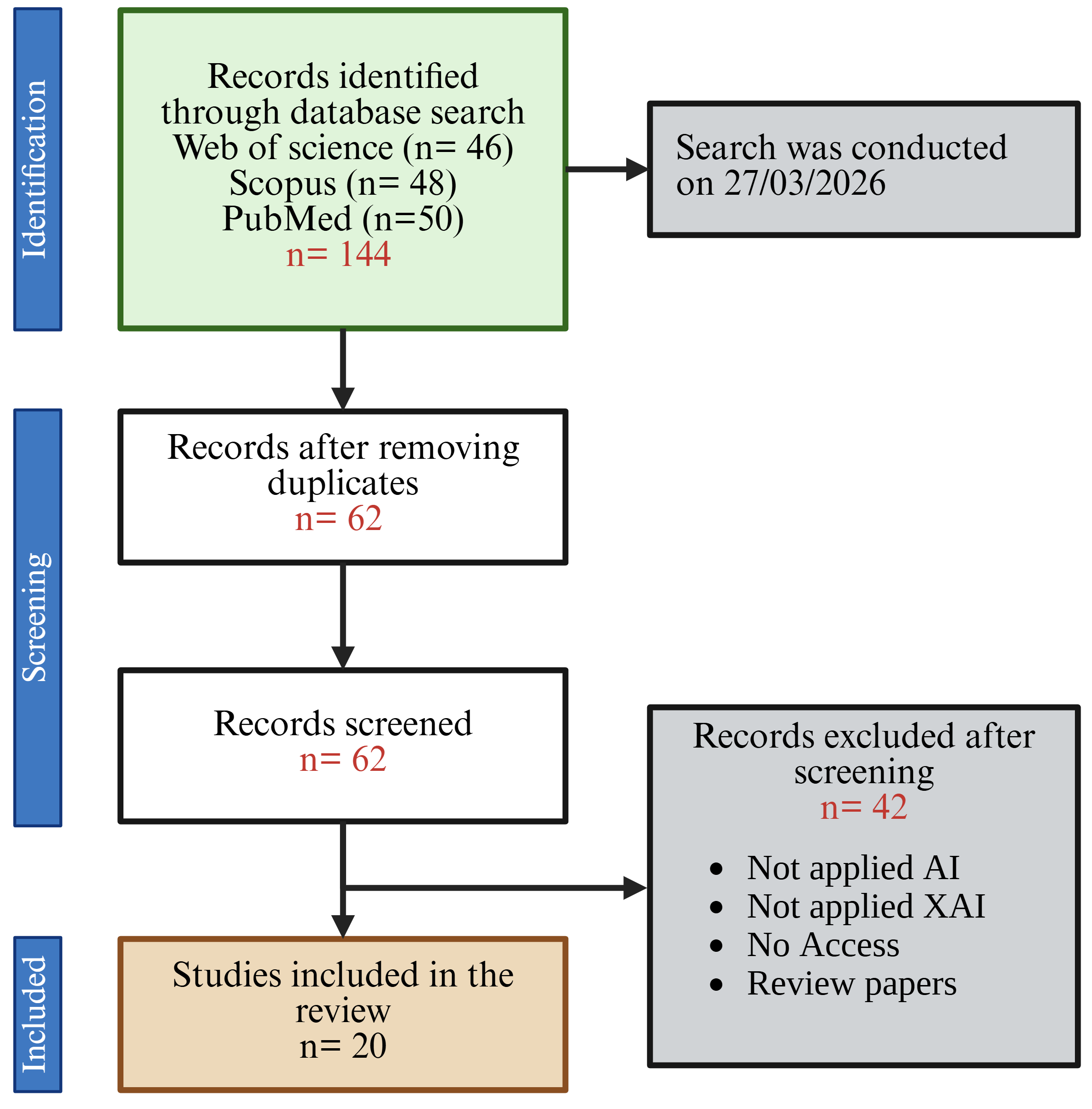}
    \caption{Flow diagram illustrating the study identification, screening, eligibility assessment, and final inclusion process for the review. The flowchart is generated by Biorender.}
    \label{flowchart}
\end{figure}

Database searching produced 144 records. After duplicate removal, 62 unique studies remained. Titles and abstracts were screened for relevance based on the eligibility criteria. A total of 42 studies were excluded for reasons including lack of AI use, absence of XAI methods, inaccessible full text, or being review papers. A total of 20 studies met all criteria and were included in the review.
The full selection process is presented in the flowchart (Figure~\ref{flowchart}). These 20 studies form the basis for the synthesis presented in this review, representing the current landscape of applying explainable AI methodologies to SBHF.
\noindent
Supplementary Table~\ref{papers} summarizes the main characteristics of the 20 studies included in this review, covering data types, clinical outcomes, reporting of sex and ethnicity, and the XAI methods used. The studies show substantial diversity in both input data and targeted cardiac outcomes, reflecting the wide range of approaches used to detect SBHF.
\subsection{Data modality}
\begin{figure}[H]
    \centering
    \includegraphics[width=0.7\linewidth]{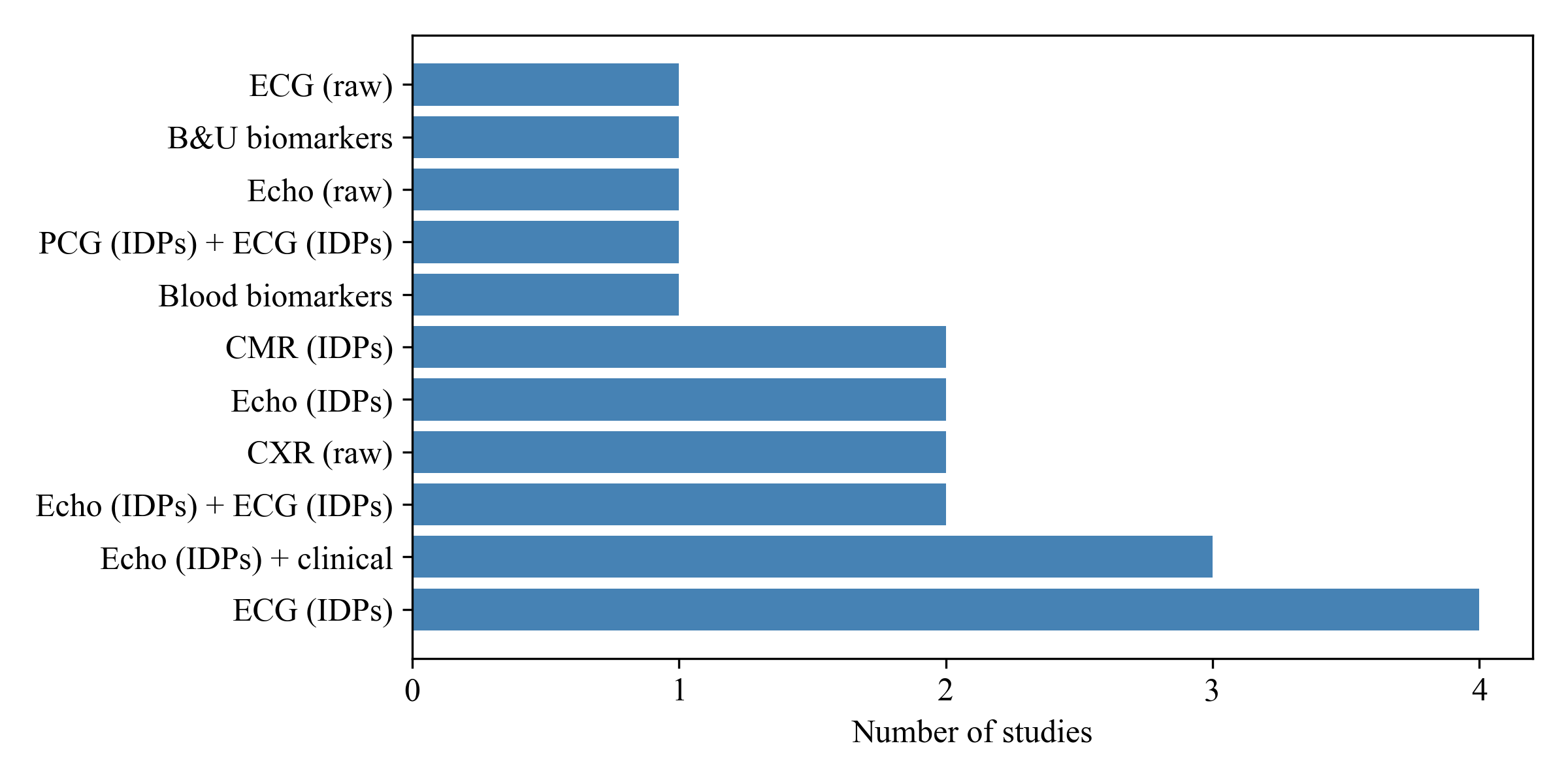}
    \caption{Distribution of data modalities used in XAI studies for stage B heart failure. Echo: echocardiography; ECG: electrocardiography; CMR: cardiac magnetic resonance; B\&U: blood and urea; PCG: Phonocardiogram; CXR: chest X-ray.  For imaging and signal data, raw indicates that data were used directly in the model, whereas IDPs (image‑derived phenotypes) refer to features extracted from images before model input. Clinical refers to patient‑level clinical variables.}
    \label{data_type}
\end{figure}

Figure~\ref{data_type} summarises the distribution of data modalities employed across XAI studies targeting SBHF. Models based on derived ECG features (ECG IDPs) are the most frequently used (n = 4), highlighting a strong reliance on extracted signal features rather than raw waveforms. This is followed by multimodal approaches combining echocardiography-derived features with clinical variables (n = 3). Other modalities, including echocardiography, cardiac MRI, and chest X‑ray, are moderately represented (each n = 2), with chest X‑ray studies in particular relying on raw image inputs. In contrast, raw data–based approaches using ECG and echocardiography remain relatively limited (mostly n = 1 per category). Biomarker-based studies and hybrid physiological signals (e.g., combined PCG and ECG) are also sparsely represented. Overall, the figure indicates a clear predominance of feature-based (IDP) approaches, alongside selective use of raw imaging data,.

\subsection{External validation}
Among the twenty reviewed studies, only four evaluated their models on external datasets~\cite{zeng2022multimodal, katsushika2023explainable, moon2025artificial, bhave2024deep}. While external testing generally resulted in reduced or comparable predictive performance, assessment of explainability generalisability was largely absent. Only one study~\cite{katsushika2023explainable} reported similar SHAP patterns across internal and external cohorts. 
\subsection{Demographics}
\begin{figure}[H]
    \centering
    \includegraphics[width=0.7\linewidth]{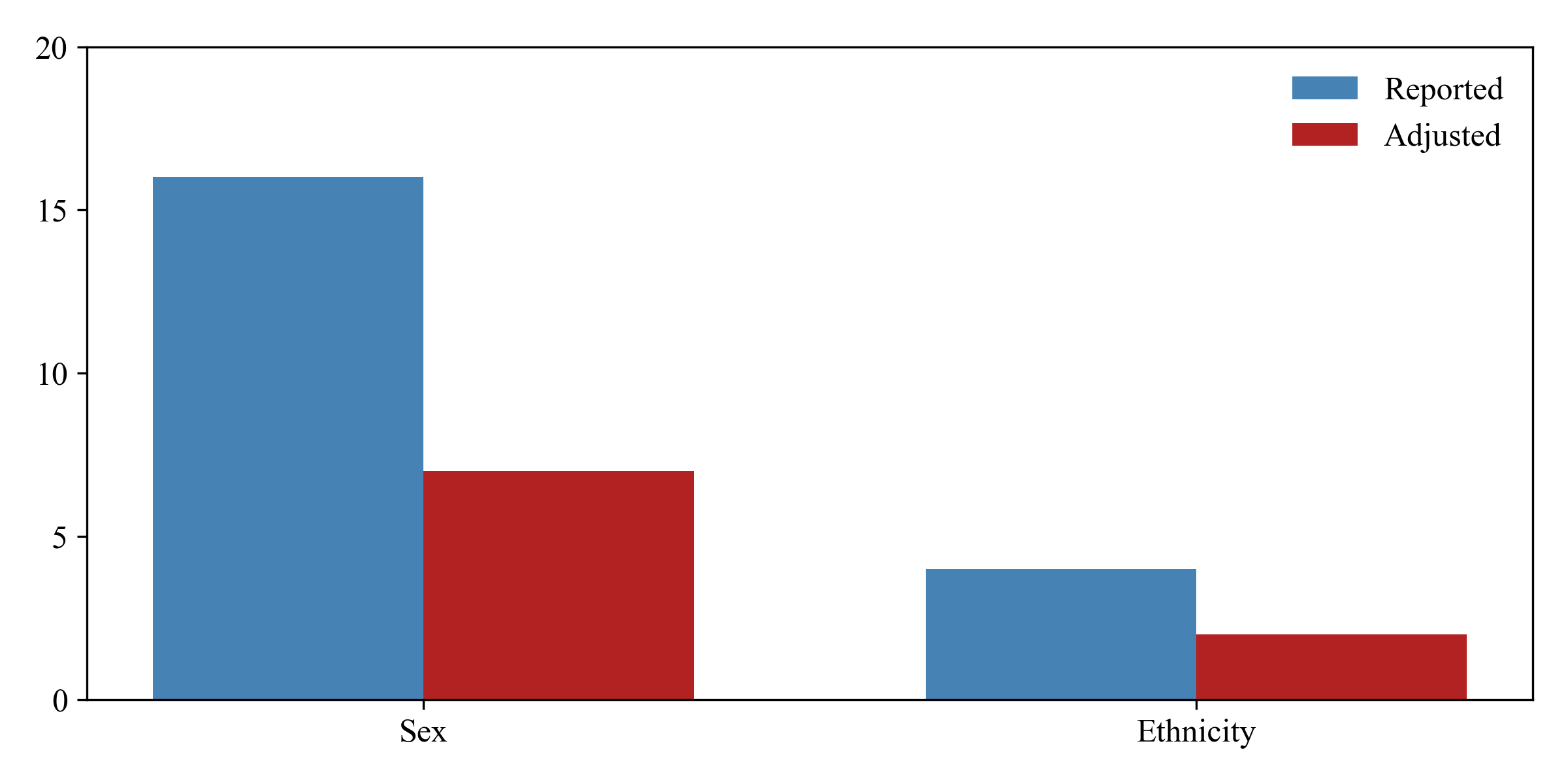}
    \caption{Reporting and adjustment for sex and ethnicity in explainable AI studies of stage B heart failure}
    \label{Demographics}
\end{figure}

Figure~\ref{Demographics} illustrates the extent to which sex and ethnicity were reported and adjusted for across the included studies. Although sex was reported in most analyses, formal adjustment for sex or ethnicity was uncommon. A small number of studies developed separate models across sex and ethnicity, with mixed findings: some reported comparable performance across subgroups~\cite{bhave2024deep,de2025clinically}, while others identified performance disparities~\cite{honarvar2022enhancing,ge2025serum}. However, none of these studies compared the generated explanations between sex or ethnic subgroups. Other studies merely included sex as an input variable~\cite{zheng2024predicting,pal2025encoding} or adjusted for sex during model development~\cite{alsharqi2023machine}.

\subsection{Outcome}
\begin{figure}[H]
    \centering
    \includegraphics[width=0.7\linewidth]{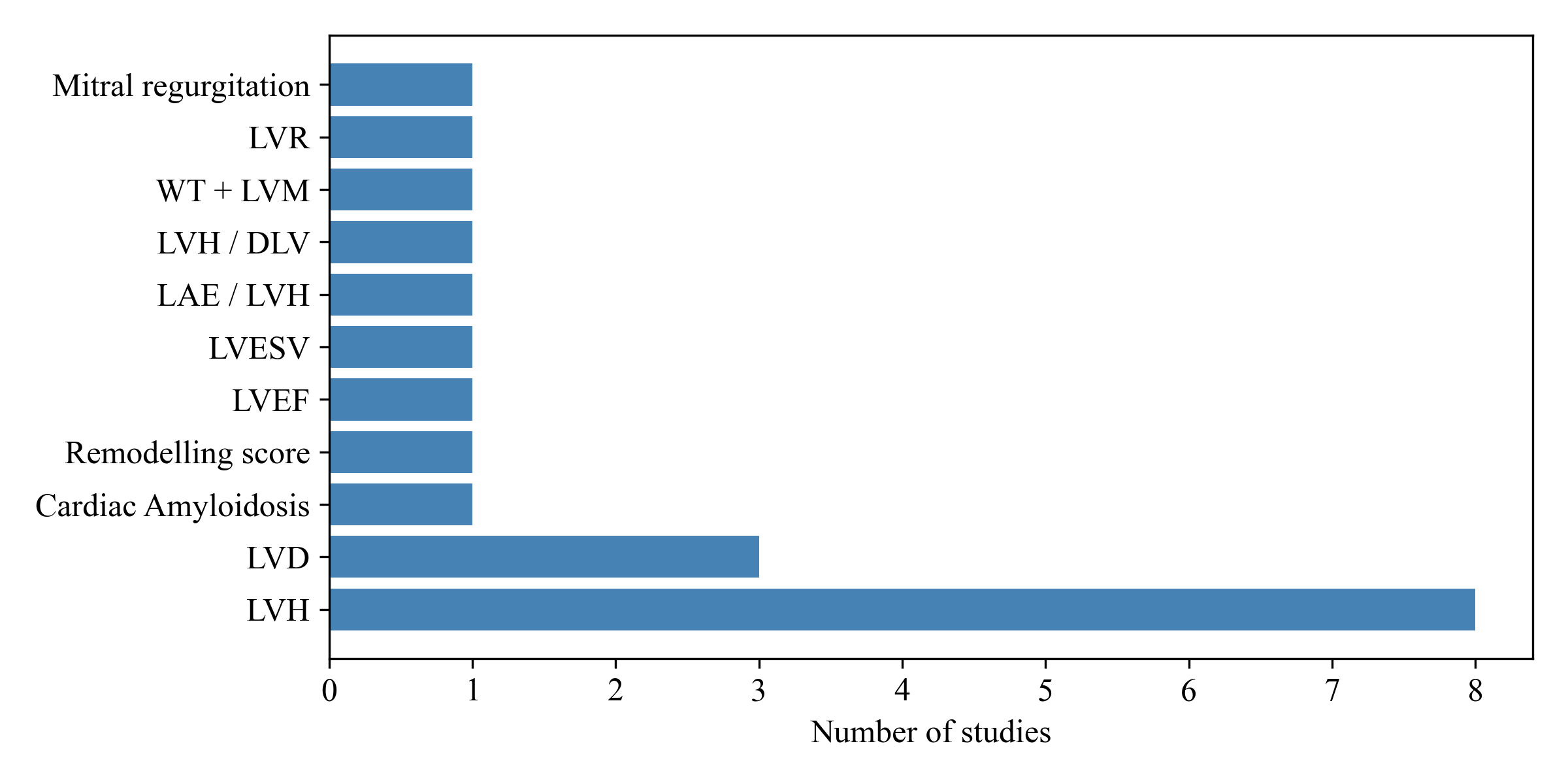}
    \caption{Heterogeneity of outcome definitions used to represent stage B heart failure. LVH: left ventricular hypertrophy; LVEF: reduced left ventricular ejection fraction; WT: wall thickness; LVESV: left ventricular end-systolic volume; LVM: left ventricular mass; DLV: dilated left ventricle; LVR: left ventricular remodelling; LVD: left ventricular dysfunction.}
    \label{Outcomes}
\end{figure}

Figure~\ref{Outcomes} demonstrates substantial heterogeneity in how SBHF is defined across studies. Although all outcomes reflect SBHF, a wide range of terms is used, including left ventricular hypertrophy, dilation, remodelling scores, volumetric indices, and condition-specific endpoints such as amyloidosis or mitral regurgitation. The predominance of LVH contrasts with many sparsely represented, conceptually overlapping outcomes. 
\subsection{XAI methods}
\begin{figure}[H]
    \centering
    \includegraphics[width=0.7\linewidth]{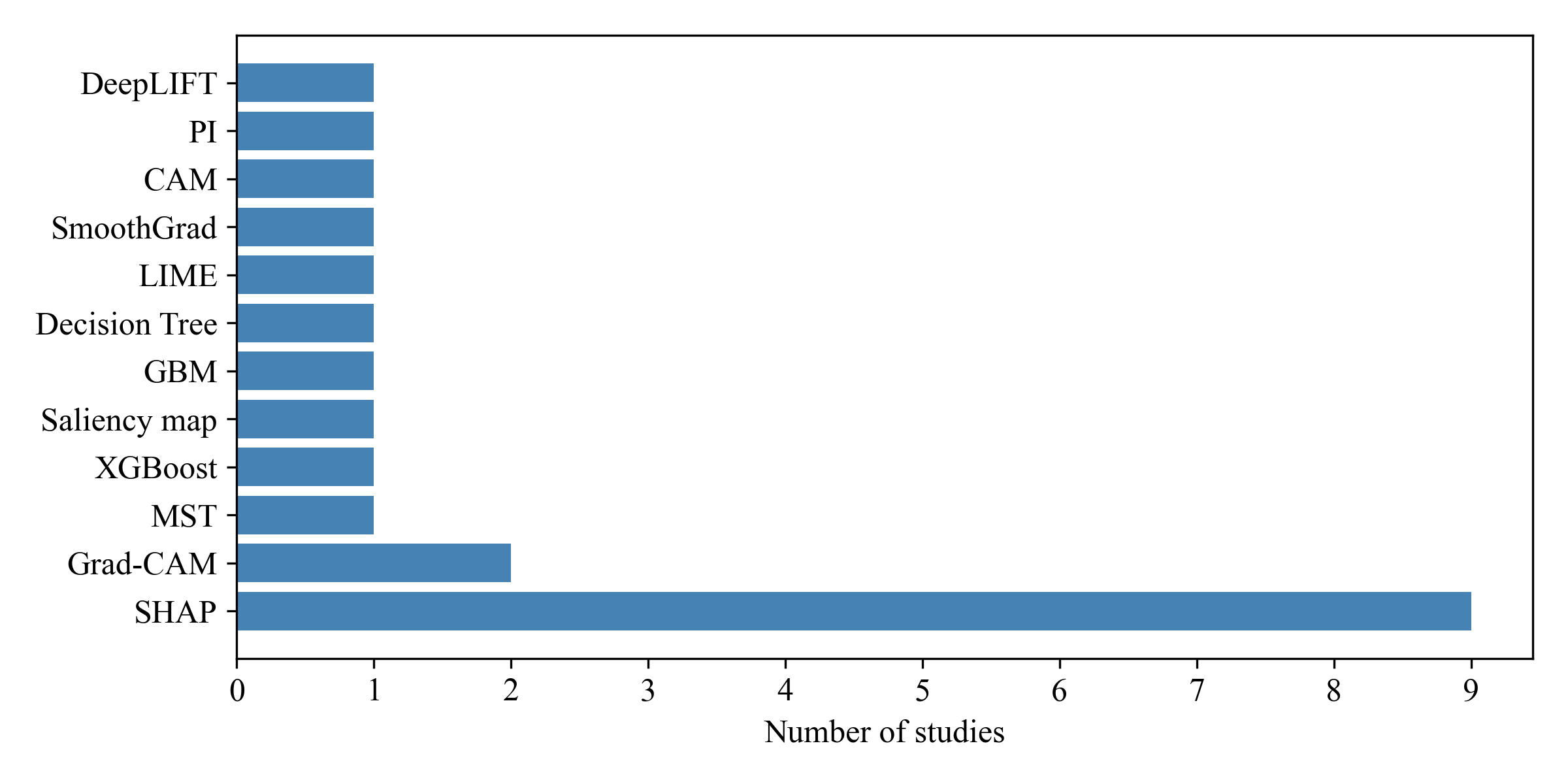}
    \caption{Distribution of explainable AI methods applied in studies of stage B heart failure. SHAP: shapley additive explanations; PI: permutation importance; MST: minimum spanning tree; LIME: local interpretable model-agnostic explanations; Grad-CAM: gradient-weighted class activation mapping.}
    \label{XAI_methods}
\end{figure}

Figure~\ref{XAI_methods} highlights a pronounced methodological skew in the application of XAI techniques. SHAP overwhelmingly dominates the landscape, reflecting a strong dependence on post-hoc feature attribution methods that are model-agnostic and relatively easy to interpret. Gradient-based visualisation methods such as Grad-CAM appear primarily in imaging-based studies but remain sparse overall. Other approaches including SmoothGrad, DeepLIFT, permutation importance, and model-intrinsic explainability are used only once, indicating limited methodological exploration. 
\subsection{XAI evaluation}
\begin{figure}[H]
    \centering
    \includegraphics[width=0.7\linewidth]{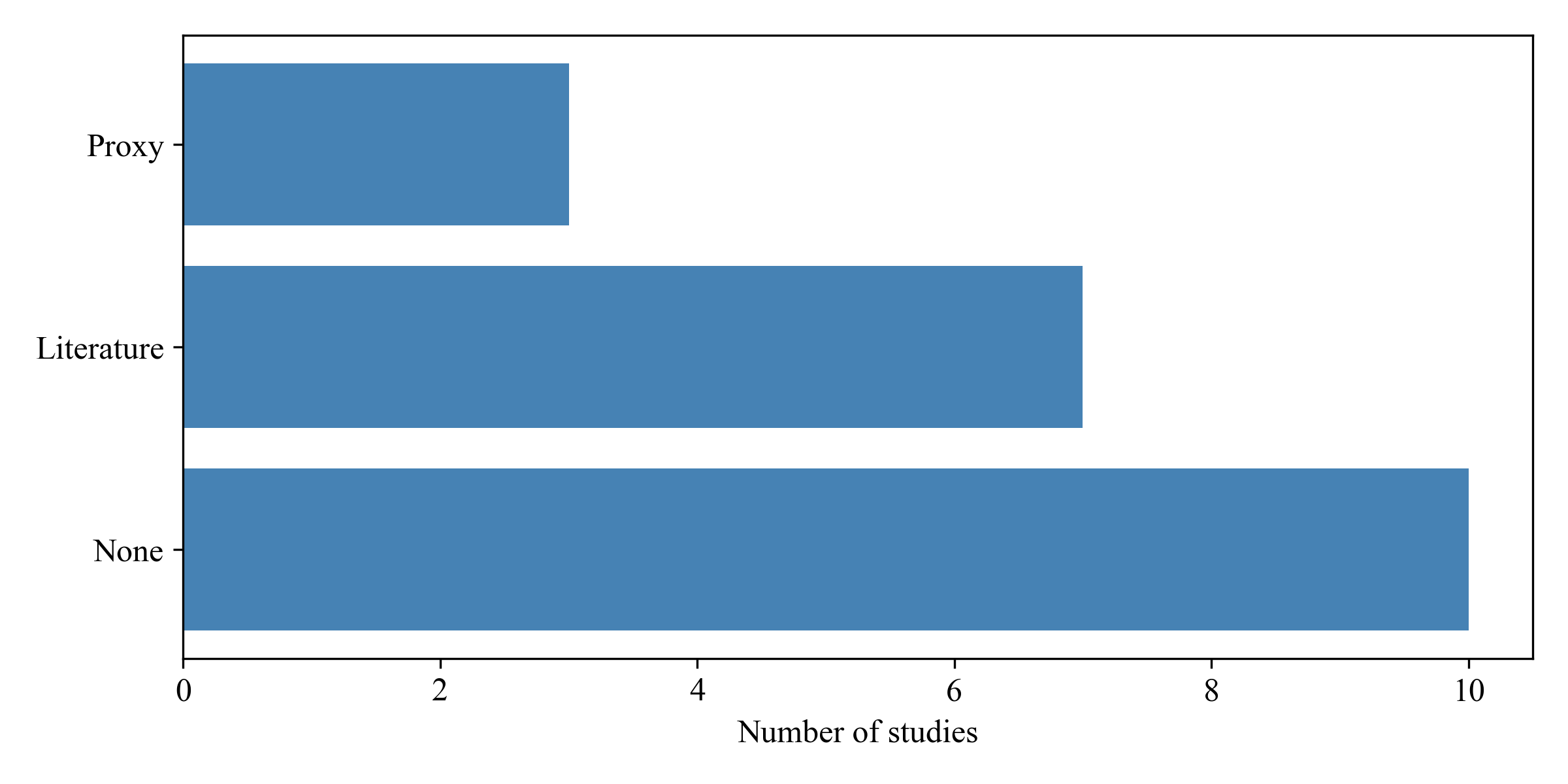}
    \caption{Evaluation strategies applied to XAI outputs in the reviewed literature.}
    \label{XAI_evaluation}
\end{figure}

Figure~\ref{XAI_evaluation} summarises the evaluation approaches used for XAI methods across the included studies. The majority of studies did not perform any formal evaluation of the XAI outputs (n = 10), indicating a substantial gap in assessing the reliability and clinical relevance of the generated explanations. Among the studies that incorporated evaluation, literature‑grounded approaches were the most common (n = 7)~\cite{yue2025interpretable, moon2025artificial, tison2019automated, angelaki2021detection, yahav2024early, zheng2024predicting, ge2025serum}, where explanations were interpreted in light of existing clinical knowledge. In contrast, proxy or functionality‑based evaluations which rely on quantitative metrics or perturbation analyses were less frequently employed (n = 3)~\cite{bhave2024deep, alsharqi2023machine, ghorbani2020deep}. \\
In their proxy‑based evaluation~\cite{bhave2024deep}, the authors used LayerCAM to assess whether the model focused on clinically relevant regions. They observed that deeper layers highlighted broad areas corresponding to the cardiac silhouette, while intermediate layers provided more localised attention within the heart, particularly the left ventricle. In contrast, shallower layers showed more variable patterns. In~\cite{alsharqi2023machine},  XAI  was evaluated via proxy-based validation: the cardiac remodelling score alignment with known pathophysiological trajectories was verified through variable-score response functions and heatmaps, and its semantic validity was tested by sensitivity to intervention-induced changes, demonstrating statistically significant associations with fitness and adherence metrics. In~\cite{ghorbani2020deep}, they evaluated XAI using saliency-based plausibility and perturbation-based faithfulness tests. Gradient-based sensitivity maps were inspected to verify that highlighted regions aligned with clinically relevant structures. Additionally, input perturbation (occlusion) was applied by masking salient regions, and significant changes in predictions confirmed that the model causally relied on those regions.

\section{Discussion}
This review provides an overview of the current use of XAI in the detection and characterisation of SBHF. Although the included studies demonstrate the growing interest in applying XAI to cardiovascular medicine, several important limitations and gaps can be identified from the results.\\
Despite the focus of this review, the application of XAI remains limited and inconsistent across studies, and even when methods such as SHAP, LIME, or Grad‑CAM are applied, their outputs are rarely rigorously evaluated. For example, SHAP assumes statistical independence among input features and does not account for collinearity, which may lead to misleading attributions~\cite{salih2024characterizing}. Notably, none of the studies involved domain experts such as cardiologists, to assess the clinical validity of explanations, only three employed quantitative evaluation approaches, and seven relied on literature‑based evaluation which is inherently biased toward prior findings that align with the generated explanations thereby limiting the reliability, objectivity, and clinical trustworthiness of XAI.\\
Second, there is a notable lack of demographic consideration across the included studies. Although sex was reported in several papers, it was not consistently adjusted for in the analysis, and ethnicity was rarely reported and adjusted for. This is a critical limitation, as pre-clinical heart failure and cardiac remodelling processes are known to vary across sex~\cite{lam2019sex} and ethnic groups~\cite{pina2021race}. Ignoring these differences may lead to biased models and reduced generalisability, ultimately limiting the robustness and clinical applicability of AI systems when deployed in diverse populations.\\
In addition, the majority of studies relied on single-modality data, most commonly echocardiography or ECG. Only a limited number of studies explored multi-modal approaches that combine different data sources. Given that SBHF is a complex and multi-factorial process involving structural, functional, electrical, and biochemical changes~\cite{cowley2026challenges}, reliance on a single data modality may restrict model performance and limit the interpretability of findings. Multi-modal AI models, combined with XAI techniques, could provide a more comprehensive understanding of disease mechanisms.\\
Another limitation relates to the heterogeneity of outcomes and definitions used across studies. While many papers focused on left ventricular hypertrophy, others targeted different but related endpoints such as left ventricular dysfunction, remodelling, or ejection fraction. This variability makes it difficult to directly compare results across studies and may hinder the development of standardized benchmarks for evaluating AI and XAI approaches in this domain.\\
External validation was implemneted in only four studies across the reviewed literature, with only one study comparing the generated explainability outcomes between the training data and an external test cohort. This lack of external validation raises concerns regarding overfitting and limits confidence in the generalisability of both the predictive models and their associated explanations. Robust validation across diverse populations, clinical settings, and data acquisition protocols is essential before such XAI approaches can be reliably translated into clinical practice.\\
In summary, while the application of XAI in SBHF is promising, the current evidence base is constrained by limited adoption of advanced explainability methods, insufficient demographic consideration, and methodological heterogeneity. Addressing these limitations will be essential for developing reliable, generalizable, and clinically meaningful AI systems for SBHF and prevention of heart failure.
\section{Open Issues and Research Directions}

Building on the findings of this review, several key limitations in the current literature highlight concrete directions for future research.\\
\textbf{1. Lack of rigorous and standardised evaluation of XAI methods} : A major gap identified in this review is the absence of systematic evaluation of XAI outputs. As shown in Figure~\ref{XAI_evaluation}, half of the included studies did not evaluate explanations at all, while others relied primarily on literature-based validation, which is inherently subjective and prone to confirmation bias. Very few studies employed quantitative-grounded evaluation. \\
Future research should prioritise the development and adoption of \textit{standardised evaluation frameworks} that combine quantitative metrics (e.g., fidelity, robustness, sensitivity), clinician-in-the-loop validation, and task-specific benchmarks. Without rigorous evaluation, it remains unclear whether the generated explanations are reliable or clinically meaningful, limiting their translational value.\\
\textbf{2. Absence of subgroup-specific explainability and fairness analysis}  Although some studies reported model performance across sex or ethnic groups, none examined whether the explanations themselves differ across subpopulations. This is a critical omission, given known variations in cardiac structure, function, and disease progression across demographic groups. \\
Future work should move beyond performance comparison and explicitly investigate whether XAI explanations are consistent or biased across subgroups, how feature importance or saliency patterns vary by sex and ethnicity, and whether such differences impact clinical decision-making. Incorporating \textit{fairness-aware XAI} will be essential to ensure equitable and generalisable AI systems.\\
\textbf{3. Limited generalisability of both models and explanations} : External validation was performed in only a small subset of studies, and one study assessed whether explanations generalise across datasets. This raises concerns that both predictions and explanations may be dataset-specific and not transferable to other clinical settings. \\
Future studies should evaluate XAI outputs across external cohorts, assess the stability of explanations under dataset shift, and investigate how differences in acquisition protocols or populations affect interpretability. Establishing the robustness of explanations is crucial for clinical trust and deployment.\\
\textbf{4. Over-reliance on single-modality and feature-based inputs} : The reviewed literature shows a strong dependence on single data modalities (e.g., ECG or echocardiography) and on derived features rather than raw data. This may limit both predictive performance and the depth of explanations, particularly for a complex, multi-factorial condition such as SBHF. \\
Future research should explore \textit{multi-modal AI models} integrating imaging, physiological signals, biomarkers, and clinical variables. Importantly, XAI methods should be adapted to provide coherent cross-modal explanations, enabling a more comprehensive understanding of disease mechanisms.\\
\textbf{5. Methodological concentration on a narrow set of XAI techniques} : The dominance of SHAP across studies indicates limited exploration of alternative explainability approaches. While SHAP is widely used, it relies on assumptions (e.g., feature independence) that may not hold in clinical data and can lead to misleading interpretations. \\
Future work should investigate a broader range of XAI methods, including model-intrinsic approaches, compare different methods systematically, and assess how methodological choices influence clinical interpretation. A more diverse methodological landscape will help identify the most reliable and context-appropriate XAI techniques.\\
\textbf{6. Unclear integration of XAI into clinical workflows} : Current studies largely focus on technical development, with minimal investigation into how clinicians interpret and use XAI outputs in practice. As a result, the real-world utility of these systems remains uncertain. \\
Future research should include human-AI interaction studies, usability and interpretability assessments with clinicians, and workflow-based evaluations in realistic clinical settings. Understanding how explanations influence clinical reasoning, trust, and decision-making is essential for safe and effective adoption.
\section*{Author Contributions}
Ahmed M Salih: conceptualization, data curation, formal analysis, methodology, project administration, validation, writing original draft.
\section*{Acknowledgments}
Figure 1 and 2 were generated using Biorender. During the preparation of the manuscript, the authors used the free plan of Chat GPT solely to assist with language clarity, flow and grammatical accuracy. All scientific content, idea and interpretations were developed, reviewed and approved entirely by the human authors, who take full responsibility for the work. The figures are generated by Biorender.
\section*{Funding}
AMS is funded by the British Heart Foundation (RE/24/130031). EMB is funded by the British Heart Foundation (RG/F/24/110125).AS is supported by the National Institute for Health and Care Research (NIHR) Biomedical Research Centre Leicester (NIHR203327) and the British Heart Foundation Centre of Research Excellence (RE/24/130031). The views expressed are those of the author(s) and not necessarily those of the NIHR or the Department of Health and Social Care.
\section*{Competing interests}
AMS and EMB declare no conflicts of interest.
\section*{Data availability}

\section*{Code availability}


\printbibliography
\newpage
\section*{Supplementary Tables}

\renewcommand{\tablename}{Supp Table}
\setcounter{table}{0}
\begin{table}[H]
\caption{Full set of search keywords related to early cardiac dysfunction and explainable artificial intelligence (XAI) used in the literature search} \label{keywords_}
\begin{tabular}{|l|l|}
\hline
\textbf{Early Heart Dysfunction Terms}                                     & \textbf{Explainable AI (XAI) Terms}                                                                         \\ \hline
Stage A heart failure                              & XAI, explainability, explainable                                                                \\ \hline
Stage B heart failure                              & Interpretability, interpretable                                                                 \\ \hline
Asymptomatic structural heart disease              & Partial dependence plots, elif                                                                  \\ \hline
Asymptomatic left ventricular dysfunction          & Accumulated local effects, activation maps                                                      \\ \hline
Left ventricular dysfunction                       & SHAPley additive explanations, SHAP                                                             \\ \hline
Preclinical heart failure                          & Deep SHAP                                                                                       \\ \hline
Pre-clinical heart failure                         & \begin{tabular}[c]{@{}l@{}}Local interpretable model-agnostic\\ explanations, LIME\end{tabular} \\ \hline
Asymptomatic left ventricular systolic dysfunction & Layer-wise relevance propagation, LRP                                                           \\ \hline
Subclinical left ventricular dysfunction           & Guided backpropagation, DeepLIFT                                                                \\ \hline
Sub-clinical left ventricular dysfunction          & SmoothGrad                                                                                      \\ \hline
Asymptomatic diastolic dysfunction                 & Saliency maps                                                                                   \\ \hline
Left ventricular hypertrophy                       & Integrated gradients                                                                            \\ \hline
Left ventricular remodeling                        & Gradient weighted class activation mapping                                                      \\ \hline
Left ventricular re-modeling                       & Grad-CAM, Grad-CAM++                                                                            \\ \hline
\end{tabular}
\end{table}

\newpage
\begin{landscape}
\begin{table}[htbp]
\caption{Characteristics of the studies included in this review, summarizing data type, key outcomes, sex, ethnicity and XAI techniques. XAI: explainable AI; PI: permutation importance; CAM: class activation mapping; GBM: Gradient Boosted Machine; MST: Minimum spanning tree; Echo: echocardiography; ECG: electrocardiogram; LGE: late gadolinium enhancement; LVH: left ventricular hypertrophy; LVEF: reduced left ventricular ejection fraction; WT: wall thickness; LVESV: left ventricular end-systolic volume; LVM: left ventricular mass; DLV: dilated left ventricle; LVR: left ventricular remodeling; LVD: left ventricular dysfunction; SHAP: shapley additive explanations; LIME: local interpretable model-agnostic explanations; Grad-CAM: Gradient-weighted Class Activation Mapping.}
\label{papers}
\begin{tabular}{|l|l|l|l|l|l|l|l|l|l|}
\hline
\textbf{Paper}                                    & \textbf{Data}                       & \textbf{\begin{tabular}[c]{@{}l@{}}External\\ validation\end{tabular}} & \textbf{Outcome}          & \textbf{\begin{tabular}[c]{@{}l@{}}Ethnicity \\ reported?\end{tabular}} & \textbf{\begin{tabular}[c]{@{}l@{}}Ethnicity \\ adjusted?\end{tabular}} & \textbf{\begin{tabular}[c]{@{}l@{}}Sex \\ reported?\end{tabular}} & \textbf{\begin{tabular}[c]{@{}l@{}}Sex \\ adjusted?\end{tabular}} & \textbf{XAI method}      & \textbf{\begin{tabular}[c]{@{}l@{}}XAI \\ evaluation\end{tabular}} \\ \hline
\cite{zeng2022multimodal}        & Phonocardiogram and ECG signals      & Yes    & LVD     & No & No  & Yes    & No   & Saliency maps    & None            \\ \hline
\cite{soto2022multimodal}        & Echo \& ECG images                  & No& LVH & Yes    & No  & Yes    & No & SHAP   & None  \\ \hline
\cite{katsushika2023explainable} & ECG and Echo features               & Yes    & LVEF    & No & No  & Yes    & No   & SHAP      & None  \\ \hline
\cite{moon2025artificial}        & Echo-based radiomics                & Yes    & LVH     & No & No  & Yes    & No   & SHAP    & Literature  \\ \hline
\cite{bhave2024deep}             & Chest X-ray images  & Yes    & Severe LVH \& DLV       & Yes  & Yes  & Yes  & Yes   & CAM & Proxy    \\ \hline
\cite{de2025clinically}          & Echo features \& clinical variables & No     & LVH     & No  & No & Yes    & Yes & Decision tree     & None   \\ \hline
\cite{honarvar2022enhancing}     & ECG features                        & No  & LVD    & Yes   & Yes & Yes & Yes & DeepLIFT     & None   \\ \hline
\cite{ge2025serum}               & Blood and urine biomarkers          & No   & LVH & No  & No & Yes    & No & SHAP & Literature  \\ \hline
\cite{zheng2024predicting}       & CMR features                        & No   & LVR & No  & No & Yes & Yes  & SHAP  & Literature \\ \hline
\cite{pal2025encoding}           & Chest X-ray images                  & No& Severe LVH     & No   & No   & Yes  & Yes  & Grad-CAM  & None  \\ \hline
\cite{alsharqi2023machine}       & Echo features \& clinical variables & No     & Cardiac remodelling score   & No   & No   & Yes    & Yes   & MST    & Proxy  \\ \hline
\cite{pan2024machine}            & Blood biomarkers    & No     & Cardiac Amyloidosis       & No   & No       & Yes  & No   & XGBoost   & None      \\ \hline
\cite{yue2025interpretable}      & LGE-scar radiomics features         & No     & LVESV   & No & No  & Yes    & No   & SHAP   & Literature   \\ \hline
\cite{tison2019automated}        & ECG features                        & No     & LVH     & Yes& No  & Yes    & No   & GBM & Literature   \\ \hline
\cite{farhad2024automated}       & Echo \& clinical variables          & No     & LVH     & No  & No & No     & No   & LIME, SHAP & None            \\ \hline
\cite{ghorbani2020deep}          & Echo images   & No & Enlarged left atrium, LVH & No    & No  & Yes   & No   & Sensitivity map      & Proxy  \\ \hline
\cite{angelaki2021detection}     & ECG & No  & WT \& LVM                 & No   & No       & Yes  & No  & SHAP    & Literature  \\ \hline
\cite{yahav2024early}            & Echo features     & No   & LVD        & No  & No   & No  & No   & PI   & Literature   \\ \hline
\cite{angelaki2026single}        & ECG single lead features            & No  & LVH     & No    & No  & No & No & SHAP & None   \\ \hline
\cite{sakuma2025utility}         & ECG images                          & No  & Mitral regurgitation      & No    & No   & No  & No   & Grad-CAM   & None   \\ \hline
\end{tabular}
\end{table}

\end{landscape}

\end{document}